\documentclass[iop,twocolumn]{emulateapj}

\usepackage{graphicx}
\usepackage{amsmath}
\usepackage{natbib}
\usepackage{mathptmx}
\usepackage{apjfonts}
\usepackage{times}
\usepackage[colorlinks=true,urlcolor=blue,citecolor=blue,linkcolor=blue]{hyperref}
\usepackage{aas_macros}
\usepackage{tikz}
\usepackage{tabularx}
\usepackage{array}
\usepackage{ulem}
\newcolumntype{L}[1]{>{\raggedright\let\newline\\\arraybackslash\hspace{0pt}}m{#1}}
\newcolumntype{C}[1]{>{\centering\let\newline\\\arraybackslash\hspace{0pt}}m{#1}}
\newcolumntype{R}[1]{>{\raggedleft\let\newline\\\arraybackslash\hspace{0pt}}m{#1}}

\bibpunct{(}{)}{;}{a}{}{,}

\usepackage[position=t,singlelinecheck=on,font={rm,bf,up},font=large]{subfig}
\RequirePackage{color}
\newcommand{\sunrise}{\textsc{Sunrise}}
\newcommand{\sufi}{\textsc{SuFI}}

\def\lesssim{\mathrel{\hbox{\rlap{\hbox{\lower4pt\hbox{$\sim$}}}\hbox{$<$}}}}

\newcommand{\caii}{Ca\,\textsc{ii}}
\newcommand{\caiih}{Ca\,\textsc{ii}\,H}
\newcommand{\caiik}{Ca\,\textsc{ii}\,K}
\newcommand{\fig}[1]{Fig.~\ref{#1}} 
\newcommand{\figand}[2]{Figs.~\ref{#1} and \ref{#2}} 
\newcommand{\tab}[1]{Table~\ref{#1}} 
\newcommand{\sref}[1]{Sect.~\ref{#1}} 

\shorttitle{Morphological properties of \caiih{} slender fibrils}
\shortauthors{Gafeira et al.}

\definecolor{darkgreen}{rgb}{0.0, 0.3, 0.0}

\newcommand{\sks}[1]{{\color{red}\textsl{!!! Sami:} #1}}

\newif\ifdraftold

\draftoldfalse

\begin{document}

\newlength{\figwidth}
\setlength{\figwidth}{9cm}%

\title{Morphological properties of slender \caiih{} fibrils observed by \sunrise~II}

\author{R.~Gafeira\hyperlink{}{\altaffilmark{1}}}
\author{A.~Lagg\hyperlink{}{\altaffilmark{1}}}
\author{S.~K.~Solanki\hyperlink{}{\altaffilmark{1,2}}}
\author{S.~Jafarzadeh\hyperlink{}{\altaffilmark{3}}}
\author{M.~van~Noort\hyperlink{}{\altaffilmark{1}}}
\author{P.~Barthol\hyperlink{}{\altaffilmark{1}}}
\author{J.~Blanco~Rodr\'{i}guez\hyperlink{}{\altaffilmark{5}}}
\author{ J.~C.~del~Toro~Iniesta\hyperlink{}{\altaffilmark{6}}}
\author{A.~Gandorfer\hyperlink{}{\altaffilmark{1}}}
\author{L.~Gizon\hyperlink{}{\altaffilmark{1}}}
\author{J.~Hirzberger\hyperlink{}{\altaffilmark{1}}}
\author{M.~Kn\"{o}lker\hyperlink{}{\altaffilmark{7}}}
\author{D.~Orozco~Su\'{a}rez\hyperlink{}{\altaffilmark{6}}}
\author{T.~L.~Riethm\"{u}ller\hyperlink{}{\altaffilmark{1}}}
\author{W.~Schmidt\hyperlink{}{\altaffilmark{8}}}

\affil{\altaffilmark{1}\hspace{0.2em}Max Planck Institute for Solar System Research, Justus-von-Liebig-Weg 3, 37077 G\"{o}ttingen, Germany; \href{mailto:gafeira@mps.mpg.de}{gafeira@mps.mpg.de}\\
\altaffilmark{2}\hspace{0.2em}School of Space Research, Kyung Hee University, Yongin, Gyeonggi 446-701, Republic of Korea\\
\altaffilmark{3}\hspace{0.2em}Institute of Theoretical Astrophysics, University of Oslo, P.O. Box 1029 Blindern, N-0315 Oslo, Norway\\
\altaffilmark{4}\hspace{0.2em}Image Processing Laboratory, University of Valencia, P.O. Box 22085, E-46980 Paterna, Valencia, Spain\\
\altaffilmark{5}\hspace{0.2em}Instituto de Astrof\'{i}sica de Andaluc\'{i}a (CSIC), Apartado de Correos 3004, E-18080 Granada, Spain\\
\altaffilmark{6}\hspace{0.2em}High Altitude Observatory, National Center for Atmospheric Research, \footnote{The National Center for Atmospheric Research is sponsored by the National Science Foundation.} P.O. Box 3000, Boulder, CO 80307-3000, USA\\
\altaffilmark{7}\hspace{0.2em}National Solar Observatory, 3665 Discovery Drive, Boulder, CO 80303, USA\\
\altaffilmark{8}\hspace{0.2em}Kiepenheuer-Institut f\"{u}r Sonnenphysik, Sch\"{o}neckstr. 6, D-79104 Freiburg, Germany
}

\begin{abstract}
We use seeing-free high spatial resolution \caiih\ data obtained by the \sunrise{} observatory to determine properties of slender fibrils in the lower solar chromosphere.
In this work we use intensity images taken with the \sufi{} instrument in the \caiih{} line during the second scientific flight of the \sunrise{} observatory to identify and track elongated bright structures. After the identification, we analyze theses structures in order to extract their morphological properties. 
We identify 598 slender \caiih{} fibrils (SCFs) with an average width of around 180\,km, a length between 500\,km and 4000\,km, an average lifetime of $\approx$400\,s, and an average curvature of 0.002\,arcsec$^{-1}$. The maximum lifetime of the SCFs within our time series of 57 minutes is $\approx$2000\,s. We discuss similarities and differences of the SCFs with other small-scale, chromospheric structures such as spicules of type I and II, or \caiik{} fibrils.
\end{abstract}


\maketitle


\section{Introduction}
\label{int}

Large parts of the solar surface are littered with small scale fibrils, loops and jets, connecting the photosphere to the chromospheric layers. These structures, seen in radiation, are thought to follow magnetic field lines \cite[e.g.,][]{jafarzadeh16d} anchored in photospheric magnetic flux concentrations, or in the weaker internetwork elements \citep{wiegelmann2010}. Except for regions with large magnetic flux concentrations, such as sunspots or large pores, these structures play an important role in transporting the energy from the solar surface to the chromosphere and to the corona, either as a channel for the propagation of waves \cite[e.g.,][]{vanBallegooijen2011}, or as the location for small-scale reconnection events \citep{Gold1964,Parker1972}. Observations in the \caii{}\,H and K lines at high spatial resolution \cite[e.g., from the Swedish Solar Telescope, SST,][]{Pietarila2009} and under seeing-free, stable conditions using the Hinode space observatory shed new light onto these small-scale structures, e.g., leading to the discovery of a new type of spicule \cite[type-II,][]{Pontieu2007b,Pereira2012}.

The unique observational conditions provided by the \sunrise{} balloon-borne solar observatory \citep{Solanki2016,Barthol2011} allow us to observe the solar chromosphere in the core of \caiih{} at constantly high temporal and spatial resolution, without the influence of seeing. This has given us the possibility to look at the structures present in the lower chromosphere at a level of detail not achieved before. Of special interest in this work are the so called slender \caiih{} fibrils (SCFs): similar to spicules or chromospheric jets, these ubiquitous features outline the magnetic field in the lower chromosphere and offer the possibility of gaining insight into the physical processes in this layer of the solar atmosphere  \citep{Pietarila2009,woger2009,jafarzadeh16b}.

In this work, we develop a technique for the automatic detection of SCFs (\sref{detmet}) enabling us to investigate their statistical properties. We discuss the lifetime, width, length, curvature, and the temporal evolution of brightenings within the individual SCFs (\sref{morph}), and compare these morphological properties to similar, small-scale structures observed with Hinode and the SST (\sref{discussion}).

\section{Data}\label{data}

The observations on which this study is based were taken by the \sunrise\ balloon-borne solar observatory \citep{Solanki2010,Barthol2011,Berkefeld2011,Gandorfer2011,Martinezpillet2011} during its second science flight \citep{Solanki2016} in June 2013, referred to as \sunrise~II. The data set used was recorded from 2013/06/12 at 23:39~UT to 2013/06/13 at 00:38~UT, and covers part of the active region NOAA 11768 including most of the following polarity, the polarity inversion line and also an emerging flux region lying between the two opposite polarity regions. At the time it was observed the active region was still young and developing and was located at  $\mu = \cos \theta = 0.93$, where $\theta$ is the heliocentric angle. The heliocentric coordinates of the center of the \sufi{} field-of-view (FOV) were $x=508$\arcsec{}, $y=-274$\arcsec{}. The one hour long time series is composed of a total of 490 images taken using the \sunrise{} Filter Imager \cite[\sufi{}, ][]{Gandorfer2011} in three wavelength bands, the \caiih{} 3968\,\AA{} line  using a narrow-brand filter with a full width at half maximum (FWHM) of 1.1\,\AA{}. The integration time of each image was 500\,ms. A broader \caiih{} channel with a FWHM of 1.8\,\AA{} recorded images with an integration time of 100\,ms, and a broad-band channel centered on 3000\,\AA{}, with a FWHM of 50\,\AA{}, delivered continuum images with an integration time of 500\,ms. The data were reconstructed using the multi-frame blind deconvolution \cite[MFBD,][]{vanNoort2005} technique to account for the image degradation by the telescope. After the MFBD reconstruction, the data have a spatial resolution close to the diffraction limit of the \sunrise{} telescope, which is at the wavelength of the \caiih{} line approximately 70\,km. The cadence (i.e. the time between two consecutive images at the same wavelength) was 7\,s. The data set, the data reduction and reconstruction are described in detail by \citet{Solanki2016}.

   \begin{figure*}
      \centerline{\includegraphics[width=\linewidth]{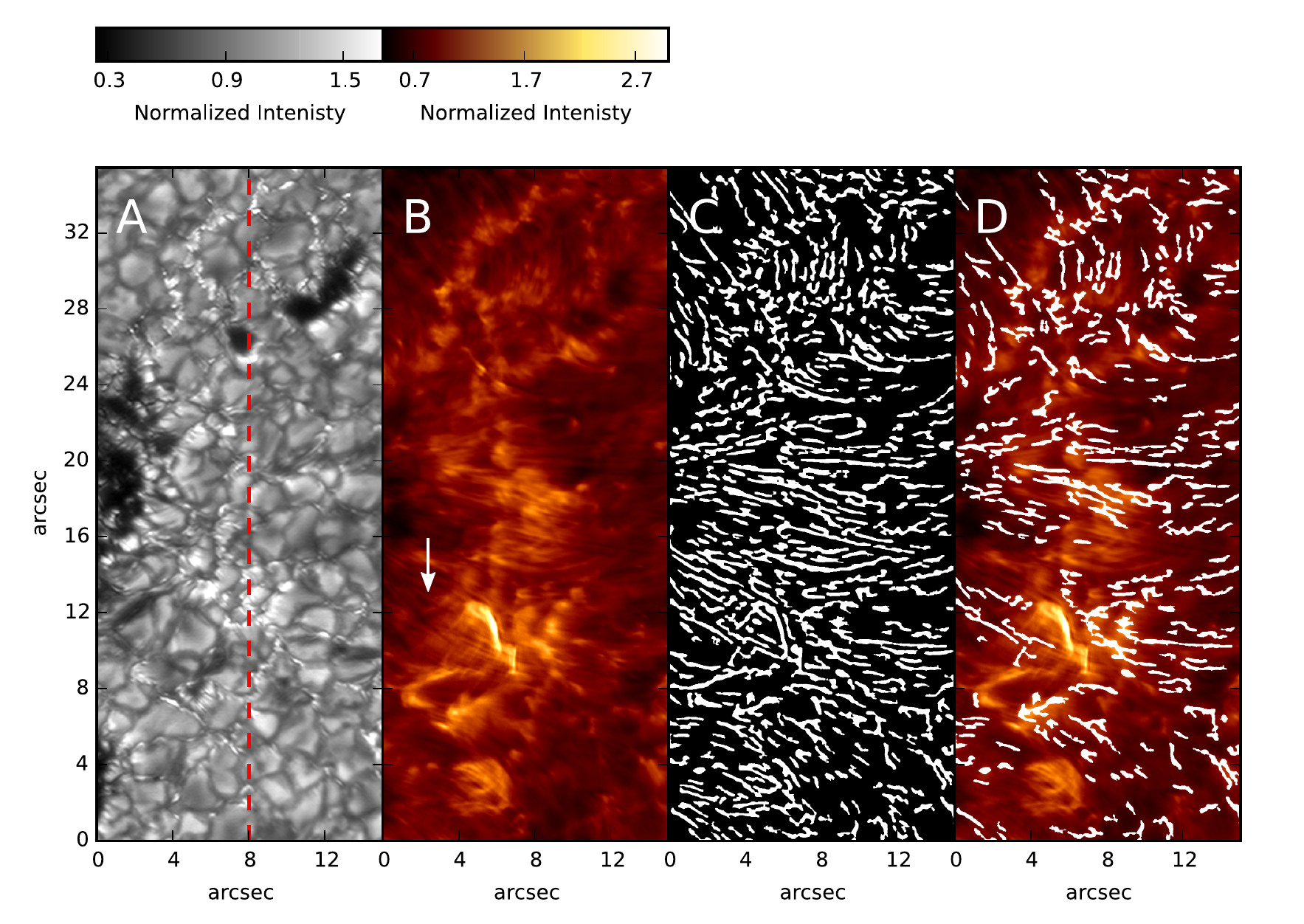}}
      \caption{(A) Continuum image recorded on 2013/06/13 at 00:31 with the 3000\,\AA{} continuum filter, (B) cotemporal, coaligned \caiih{} image, (C) binary mask resulting from the fibril identification method, and (D) superposition of the \caiih{} image and the identified fibrils. The dashed red line on image A indicates the position of the slit used in \fig{spacetime}. The white arrow on image B points to a sample fibril that is shown in greater detail in \fig{fibilexample}.}
         \label{allims}
   \end{figure*}

\section{Fibril detection and tracking}\label{detmet}

The main goal of this work is to determine the basic morphological properties of the SCFs. To this end we must first identify and track the bright elongated structures that can be seen in \fig{allims} B. To perform this identification we apply a series of image processing and contrast enhancement techniques. 
We start by subtracting a boxcar-averaged version of the original image from itself to remove the low frequencies and to enhance the structure with the typical dimensions of the fibrils. The size of the boxcar window was set to 20 pixels, corresponding to approximately 0.4\arcsec{}. Then, we apply a sharpening filter using the UNSHARP\_MASK function from the Interactive Data Language (IDL, Exelis Visual Information Solutions, Boulder, Colorado) to increase the global image contrast. This unsharp mask involves the following steps: (i) the original image is smoothed with a Gaussian filter having a width of 10 pixels; (ii) this smoothed image is then subtracted from the original image; and (iii) the resulting difference image is again added to the original image. In a next step, an adaptive histogram equalization is performed, described in detail by \citet{Pizer1987}, again using the implementation in IDL (version 8.3) with the standard parameter settings to further increase the contrast. Finally, we apply a boxcar smoothing with a width of 3 pixels to remove frequencies beyond the spatial resolution of \sunrise{}, introduced by the steps described above.

The resulting contrast-enhanced images highlight most of the fibrils very well  and allow the application of a binary mask with a threshold of 50\% of its maximum intensity,  separating the fibrils from their surroundings. The same threshold was used for the entire field of view of \sufi{} and for all frames in the time series, ensuring an unbiased determination of the length, width and shape of the identified fibrils.

Finally, we exclude all detected regions smaller than 200 pixels in area (corresponding to $\approx$0.1~arcseconds$^2$), which is approximately the area of a circle corresponding to a spatial resolution element close to the diffraction limit. 
An example of such a binary mask resulting from this process is shown in \fig{allims} C.

The next step is to track individual fibrils in time to analyze their evolution. For this task we take 6 consecutive frames and identify those regions where the binary mask marks the presence of a fibril in at least 10 pixels at the same image position in at least 5 out of a total of 6 consecutive frames. After this, we assign each fibril a tracking index. Then we advance one frame and repeat the same analysis, checking whether each newly identified fibril already has a tracking index. If yes, it is counted as the temporal continuation of the fibril from the previous frame; if not, it is counted as a new fibril.

This method ensures, with a high level of confidence, that we catch the temporal evolution of individual fibrils. The method works very well for nicely separated fibrils, but it has some limitations in two special cases: i) if a significant number of other fibrils cross the identified fibril, the binary mask will show either interruptions or the crossing fibrils are counted towards the identified fibril, and ii) if the identified fibril has a high transverse velocity (i.e., perpendicular to the fibril axis), it will move out of the detection window in the next frame.
The latter effect is particularly important in the regions of magnetic flux emergence (near $x=12$\arcsec{}, $y=9$\arcsec{}) where the fibrils are much more dynamic than in the rest of the \sufi{} FOV.

Due to the large density of fibrils, about 80\% of the detected fibrils suffer from the presence of crossing or very close neighboring fibrils. The identified fibril and the fibrils crossing it are often misidentified as a single structure with a complex, frayed shape instead of the expected linear, elongated one identified by the white arrow in \fig{allims}. If such a complex shape does exist over several consecutive frames, a correct fibril identification using a binary mask is nearly impossible. 

\begin{figure}[ht]
   \centering
      \includegraphics[width=\figwidth]{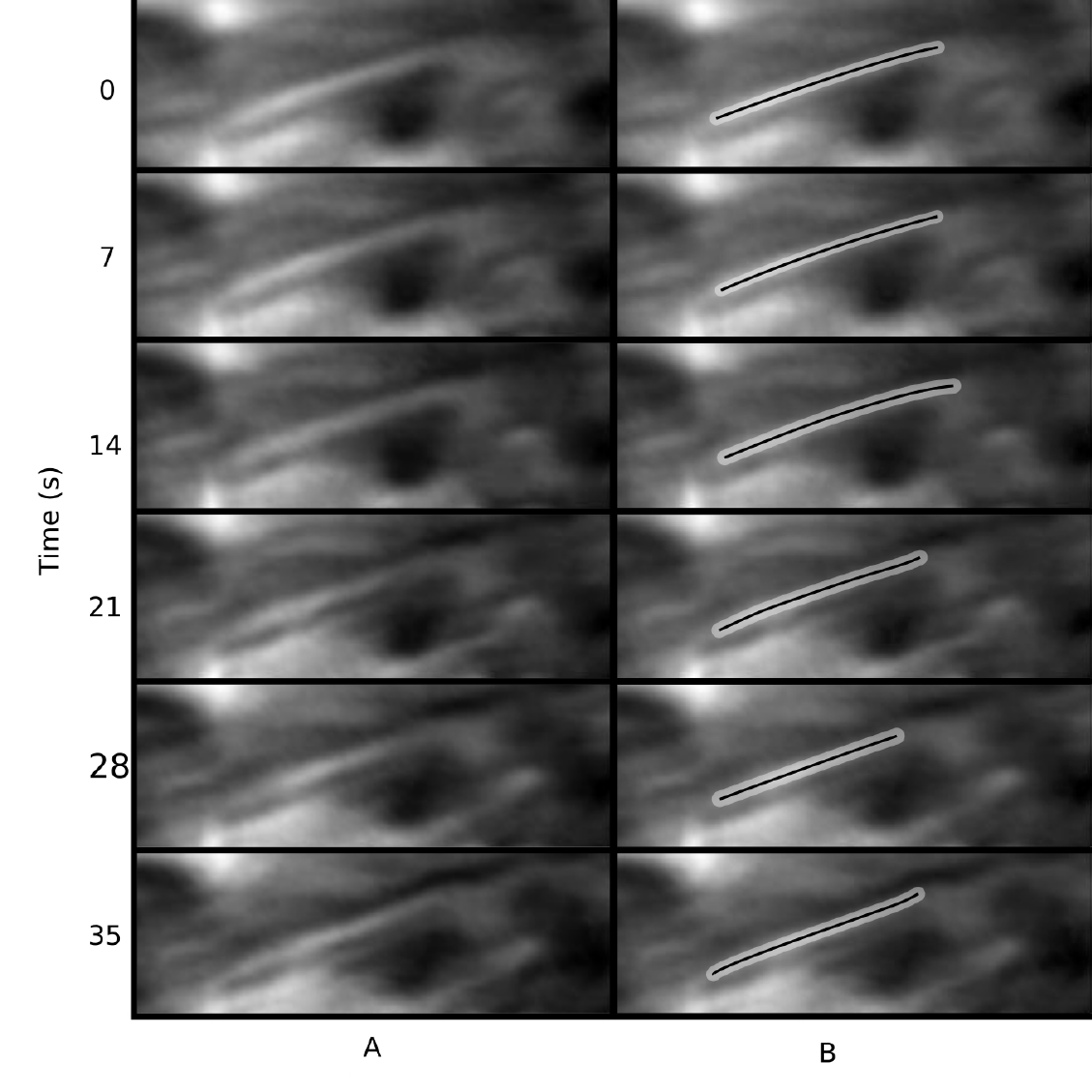}
      \caption{Example of the temporal evolution of a tracked fibril. The left panels show the original \caiih{} images, representing the temporal evolution of a fibril identified over a set of six consecutive \sufi{} frames. The right panels show the same images, overplotted with the result of the detection and tracking method described in the main text. The black lines represent the individual backbones and the gray regions indicate the width of the fibril computed using the method explained in \sref{Widths}.}
         \label{fibilexample}
   \end{figure}

To address this problem we introduce the concept of a fibril's reference backbone, which we define as the second order polynomial, best fitting the long axis of the fibril and therefore marking the ridge of maximum brightness. It is computed by first reducing the fibril in every frame of the time series of \caiih{} images to a shape that is a single pixel in width. Each such pixel is equidistant from the fibril's lateral boundaries (which are the locations where the brightness drops below the threshold).
The set of all such (single) pixels constitute the fibril's backbone. Afterwards, the fibrils of the individual frames are fitted with a second order polynomial. This polynomial is extended by the average width of this fibril (see \sref{Widths}), on both end points of the fibril to compensate for the shortening due to the reduction of the fibril to a single pixel line. The resulting curve we define as the reference backbone.
Fibrils of complex shape, often produced by intersecting fibrils or by not well separated fibrils, are usually poorly fitted ($\chi^2$ values are higher than 6), and are excluded from this averaging.
An example of this process is illustrated in \fig{fibilexample}, where we show one identified fibril and its temporal evolution (left panels) and the overplotted  fibril backbone (black line, right panels). The gray shaded area around the skeleton displays the width of the fibril, computed in every single frame using the method described in\sref{Widths}.

The fibril is well identified in all six plotted time steps. However, its length may well be underestimated, as the detection method often fails to follow the fibril to its very end, as can be seen by eye. Fibrils often become too faint near their ends to be reliably identified by our method.
Being faint, they are also susceptible to interference from background structure and crossings by other fibrils. We have therefore not attempted to extend the lengths of the individual fibrils. Instead, we consider that the fibrils that we have identified are often in fact fibril  fragments \citep[cf.][]{Pietarila2009}, i.e., parts of longer fibrils. For brevity we will continue to call them fibrils (or SCFs) in the following, although in many cases we may be discussing only parts of the full fibrils. 

We obtain a total of 598 fibrils in the \sufi{} FOV over the entire time series. 
One snapshot with the identified fibrils overlayed on top of the corresponding \caiih{} image is presented in panel D of \fig{allims}.

\section{Fibril morphology}\label{morph}

The result of the fibril tracking allows for a statistical analysis of the SCFs and to compute several of their morphological properties. In this section we present the results obtained for lifetime, length, width, and curvature of SCFs extracted using information obtained from the detection and tracking methods.

\subsection{Lifetime}\label{sec:lifetime}

Visual inspection of the movie created from the time series of \sufi{} images (see online material, \texttt{sufi.mp4}) reveals the highly dynamical nature of the SCFs. The identification of the individual fibrils over multiple time steps allows producing a lifetime histogram with a lower limit of 35 seconds (corresponding to 5 consecutive frames, which is the minimum length of time we require for a fibril to survive to be identified as such) and an upper limit given by the length of the time series (57 minutes). Only fibrils with a defined start- and end-time are included in the histogram, i.e., fibrils that are observed in the first or the last frame of the series are excluded from this analysis. 

\begin{figure}[ht]
  \centering
  \includegraphics[width=\figwidth]{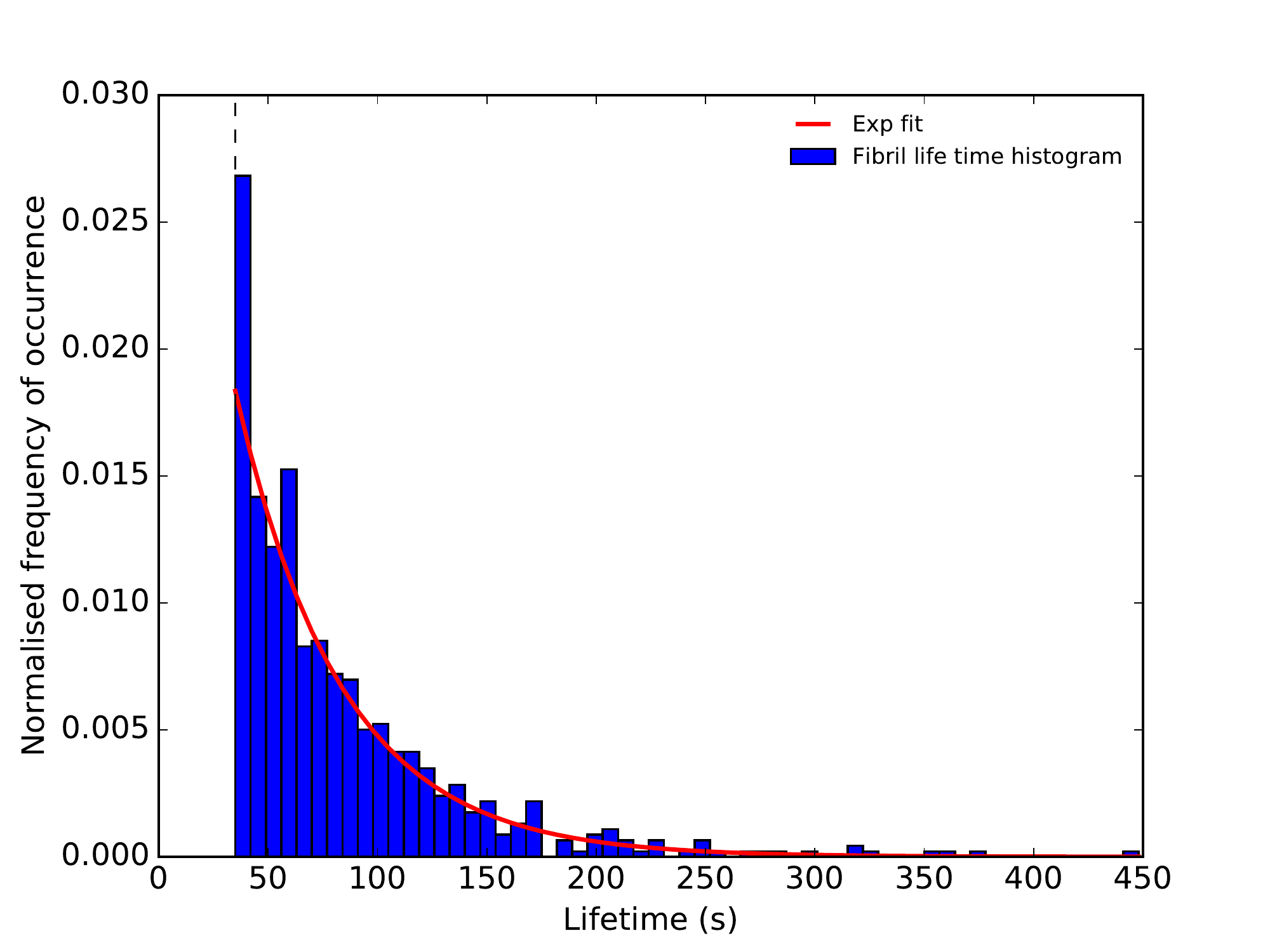}
  \caption{Fibril lifetime distribution and an exponential fit to the observed distribution (red line). Plotted is the frequency of occurrence of fibrils per lifetime bin versus the lifetime. Only fibrils with a minimum lifetime of 35\,s are included.
  }
  \label{lifetime}
\end{figure}

This lifetime histogram is presented in \fig{lifetime} (all histograms shown in the present paper were normalized so that they have a total area equal to 1). It shows a clear exponential decay with increasing lifetime at a decay rate of 2.5$\times 10^{-2}$\,$s^{-1}$ (red line). About 80.3\% of the fibrils have lifetimes between 35\,s (i.e., the lower threshold) and 100\,s, and only 1.6\% live longer than 300\,s (see also \tab{morphprop}). It must be noted that these lifetimes are likely underestimated for multiple reasons. By considering only fibrils that are born and die in the course of the observed time series, we bias the histogram towards shorter-lived fibrils due to the finite length of the time series. Changes in either the fibril or its background or neighboring fibrils can also cause its lifetime to be underestimated. Changes in the brightness or position of the fibril can cause the tracking algorithm to lose it, thus curtailing its apparent lifetime. Similarly, variations in the overall brightness and/or quality of the image may cause an apparent disappearance and reappearance of the same fibril, which is then counted as two individual fibrils, both of them with a shorter apparent lifetime. In addition, fibrils well visible on the dark background of the outer parts of the pore and its immediate surroundings become hard to follow above the bright magnetic concentrations in the right part of the image. Also, there appear to be weak, short and strongly inclined fibrils connecting bipolar regions in the right part of the figure that are directed from the NE to the SW, which cross the longer mainly E-W directed fibrils. Similarly, in the upper half of the figure, fibrils emanating from the large pore often cross other fibrils pointing in different directions. Hence the automated identification catches only individual relatively undisturbed fragments of fibrils.

   \begin{figure}[ht]
   \centering
	   \includegraphics[width=\figwidth]{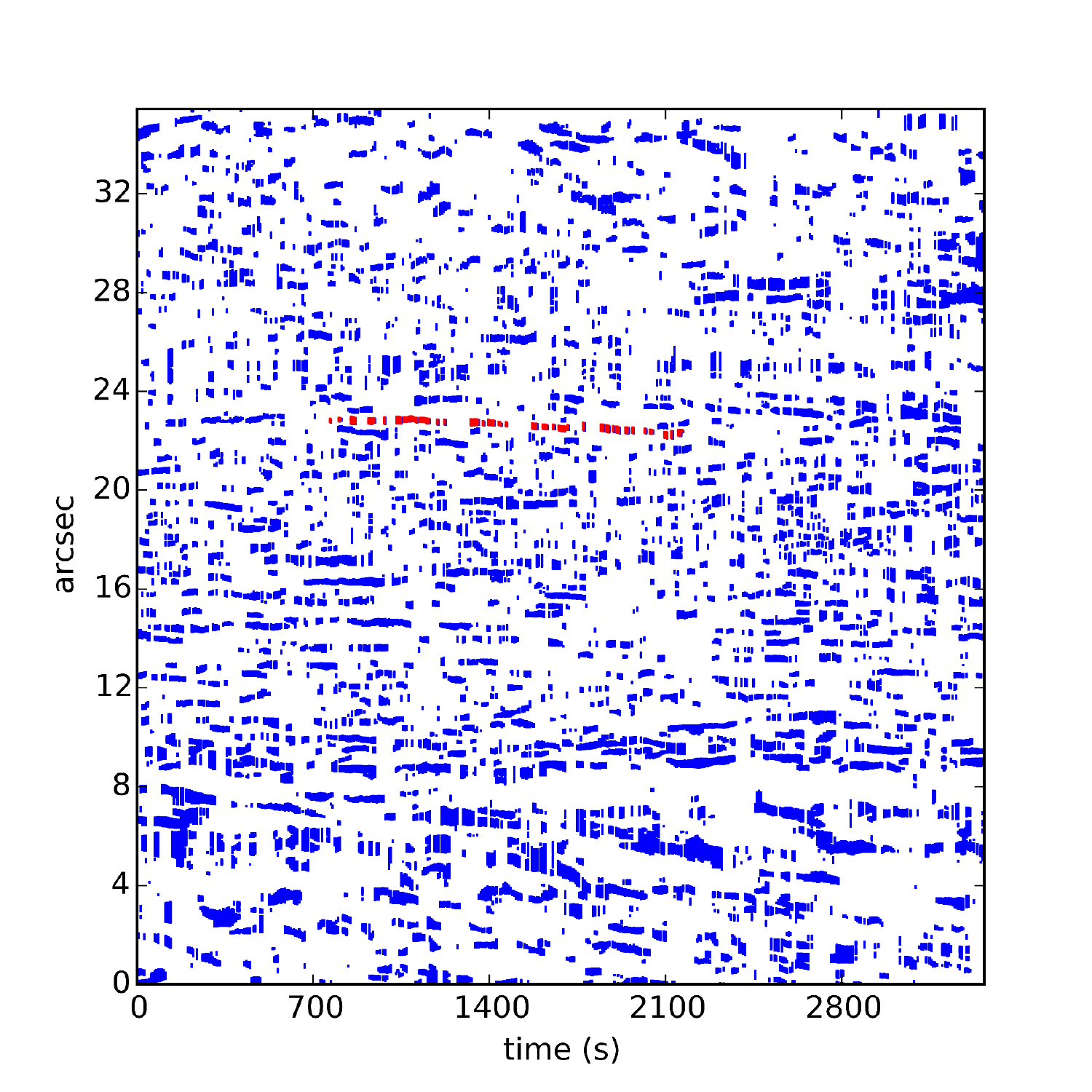}
   \caption{Space-time diagram of the presence of fibrils along an artificial slit indicated by the red, dashed line in \fig{allims} A.
   One fibril is marked in red to exemplify its intermittent appearance.}
   \label{spacetime}
   \end{figure}
   
To investigate this effect, introduced by the tracking method, we analyzed the position of the fibrils in a space-time diagram. \fig{spacetime} shows the temporal evolution of fibrils along an artificial slit at a fixed $x$-position, $x=8$\arcsec{}, indicated by the red, dashed line in \fig{allims} A. The slit stays fixed at this position throughout the whole time series. Plotted is a binary mask, with the presence of (identified) fibrils indicated in blue. In many cases, the blue patches look like an intermittent, nearly horizontal line, indicating that this particular fibril first disappears and then reappears at almost the same location. For better visibility, one such intermittent line is highlighted in red.
Most of the longer interruptions indicate a true darkening (or complete disappearance) of the fibril, with a later reappearance within the same magnetic flux tube or set of field lines, which obviously has a lifetime longer than the fibril itself. Also, some of the brightness variations leading to interruptions of fibrils in a space-time plot may be related to the sausage-mode waves identified in the fibrils by \citet{gafeira16b}.

To identify the intermittent fibrils, exemplified by the one marked in red in the space-time diagram presented in \fig{spacetime}, we use exactly the same approach as described in \sref{detmet}, but with different parameter settings, allowing to bridge the longer temporal gaps. In \sref{detmet}, we required that 10 common pixels must exist in at least 5 out of 6 consecutive frames, to ensure that we are really tracing the same fibril. Here, we relax this condition and require that 10 common pixels must exist in only 2 out of 16 frames. This closes most of the temporal gaps visible in \fig{spacetime}, at the expense of an increased likelihood of catching another fibril, which moved into the same position as the previous fibril within $\leq$105\,s (15 frames).

\begin{figure}[ht]
   \centering
	   \includegraphics[width=\figwidth]{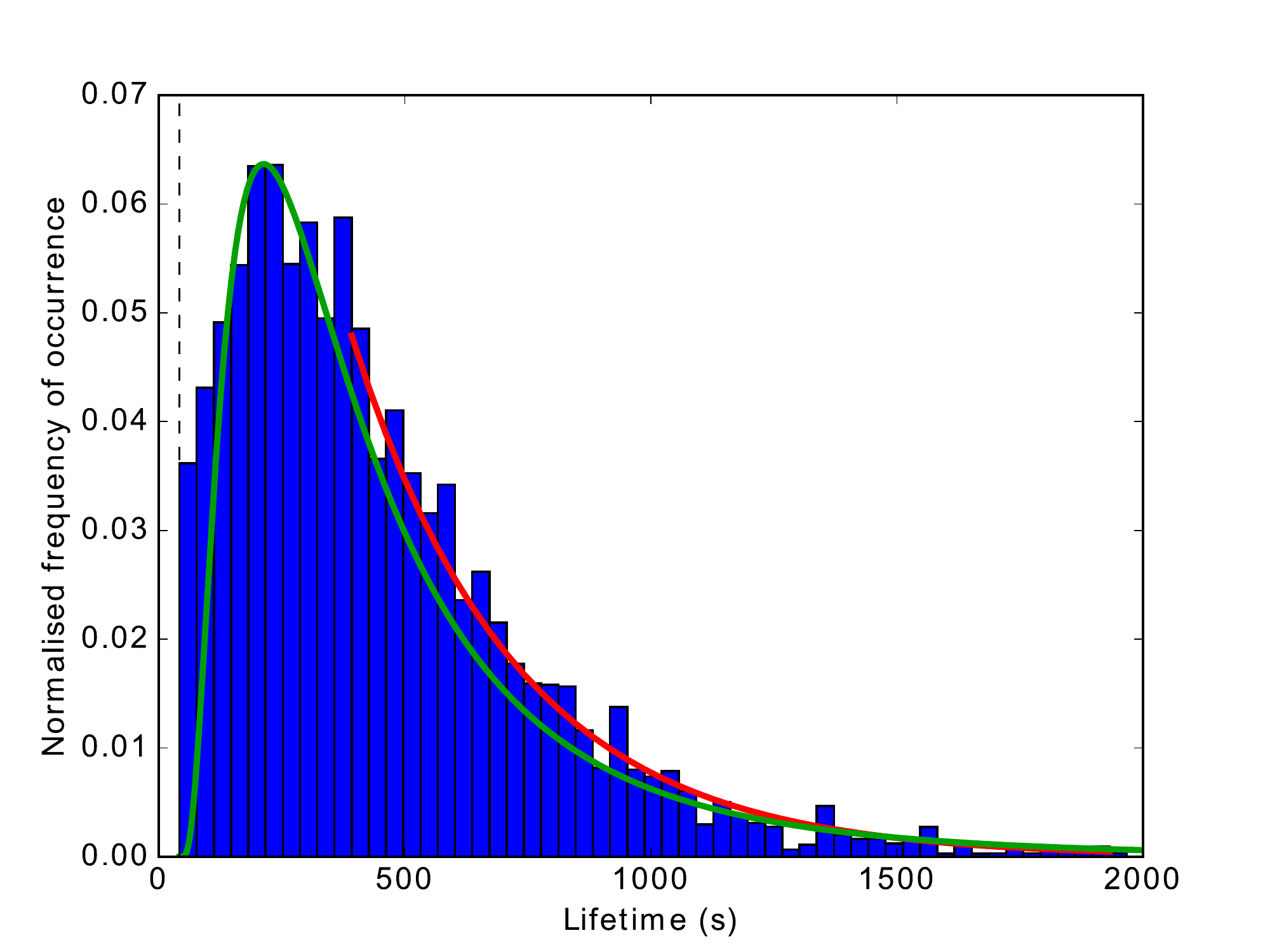}
   \caption{Fibril lifetime distribution taking into account the intermittency of the detected fibrils. The red line indicates an exponential fit for fibrils with a lifetime $\ge400$\,s. The red line indicates the respective log-normal fit.
}
   \label{fig:extendlife}
   \end{figure}

The so determined lifetime is still underestimated due to the maximum allowed
gap length of 15 frames, and the finite length of the time series, as
longer-lived fibrils are more likely excluded because they were alive
either at the beginning or the end of the time series. The latter effect can be corrected following \cite{Danilovic2010}: we multiply the frequency of
occurrence of fibrils that live for $m$ frames by the factor $(n-2)/(n-1-m)$,
where $n$ is the total number of frames in the time series. This correction has been taken into account in \fig{fig:extendlife}. The average lifetime of the fibrils obtained after all these corrections is \fig{fig:extendlife} is 446\,s.
Other parameters describing the lifetime distribution are listed in \tab{morphprop}.

We then compared this lifetime distribution with the one from the automatic tracking method. The new distribution, presented in \fig{fig:extendlife}, now shows a maximum at a lifetime of $\approx$150--300\,s. More than 58\% of the fibrils have lifetimes longer than 300\,s, 8.5\% even live longer than 15 minutes. The red line indicates an exponential fit to the tail of the distribution (i.e., to the distribution of lifetimes longer than 400\,s) and the red line indicates a log-normal fit to the entire distribution. The decay rate is now found to be $3.0\times 10^{-3}$\,s$^{-1}$, meaning that the distribution drops off nearly an order of magnitude slower than before the correction of the lifetimes for the intermittency. Obviously, the magnetic structure (possibly some kind of flux tube) housing the SCFs is considerably longer lived than the brightenings in the \caiih{} line. It could also be that the source of the brightening (located, e.g., in the photosphere) is intermittent, causing a bright fibril to appear equally intermittently.

\subsection{Width}\label{Widths}

\begin{figure}[ht]
\centering
\includegraphics[width=\figwidth]{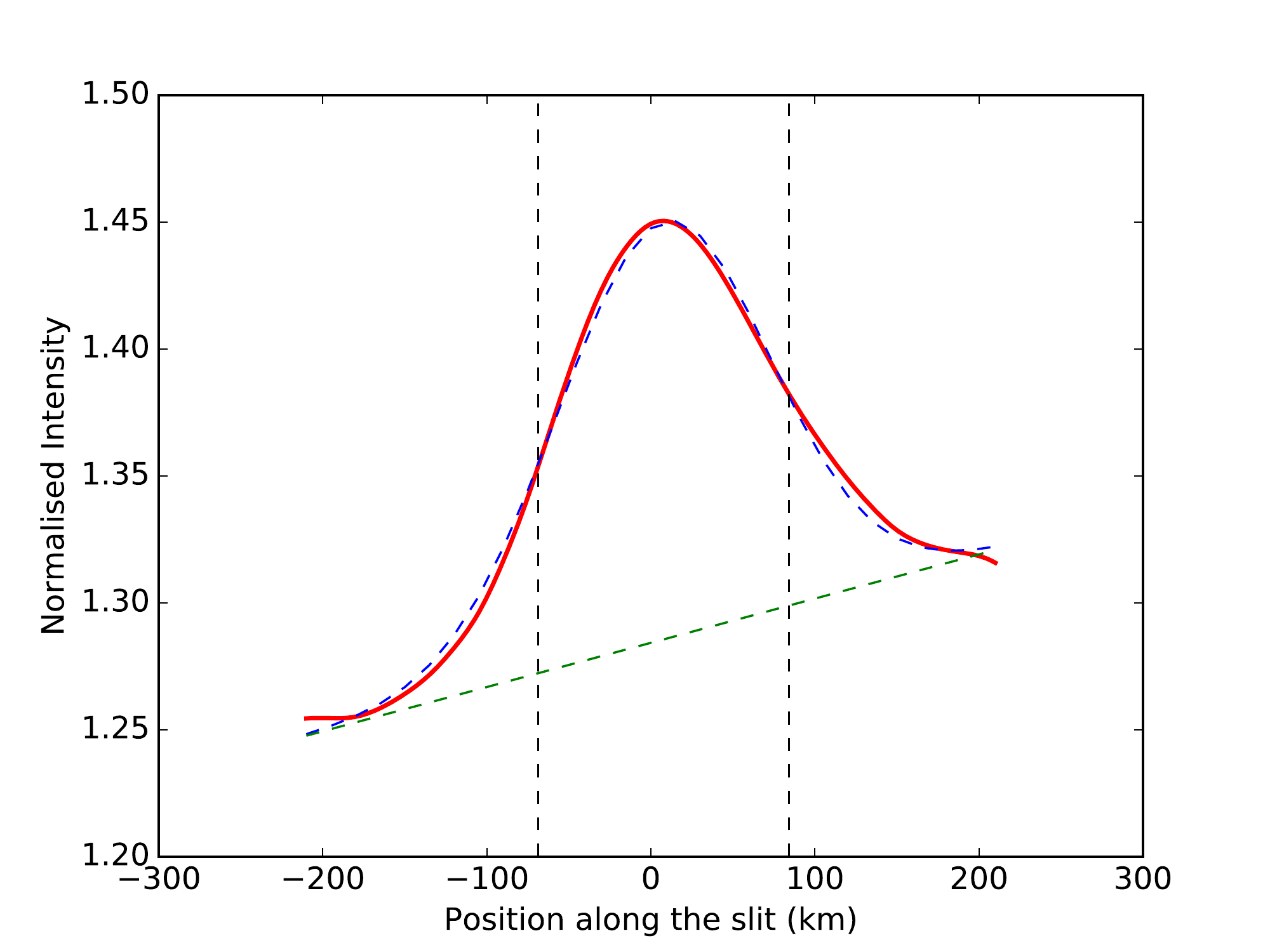}
\caption{Example for the determination of the fibril width by fitting a superposition of a Gaussian and a linear function (dashed blue line) to the brightness variation (red line) perpendicular to the fibril backbone. The linear function (dashed green line)  takes into account the variation of the background brightness. The width of the fibril, indicated by the vertical dashed lines, is the FWHM of the Gaussian.}
\label{exwidth}
\end{figure}

The automatic detection method delivering the backbone of the fibrils facilitates the computation of the widths of the SCFs. This computation, illustrated in \fig{exwidth}, is done by first selecting the position of the maximum intensity of a fibril along a virtual slit, with a length of 0.6\arcsec{}, oriented perpendicularly to the backbone of the fibril and placed at the center of the fibril's backbone. Then we fit a weighted Gaussian function plus a linear background to the intensity on the slit. Points closer to the maximum intensity are given larger weights, in order to avoid that the fit is negatively influenced by close neighboring fibrils. We then take the full width at half maximum (FWHM) of the Gaussian as a measure of the fibril's width. To further minimize the influence of neighboring or intersecting fibrils we require that the side flanks of the fibril to be well visible and the linear background brightness variation to have a slope smaller than 0.002 per km in units of the normalized intensity.

\begin{figure}[ht]
\centering
\includegraphics[width=\figwidth]{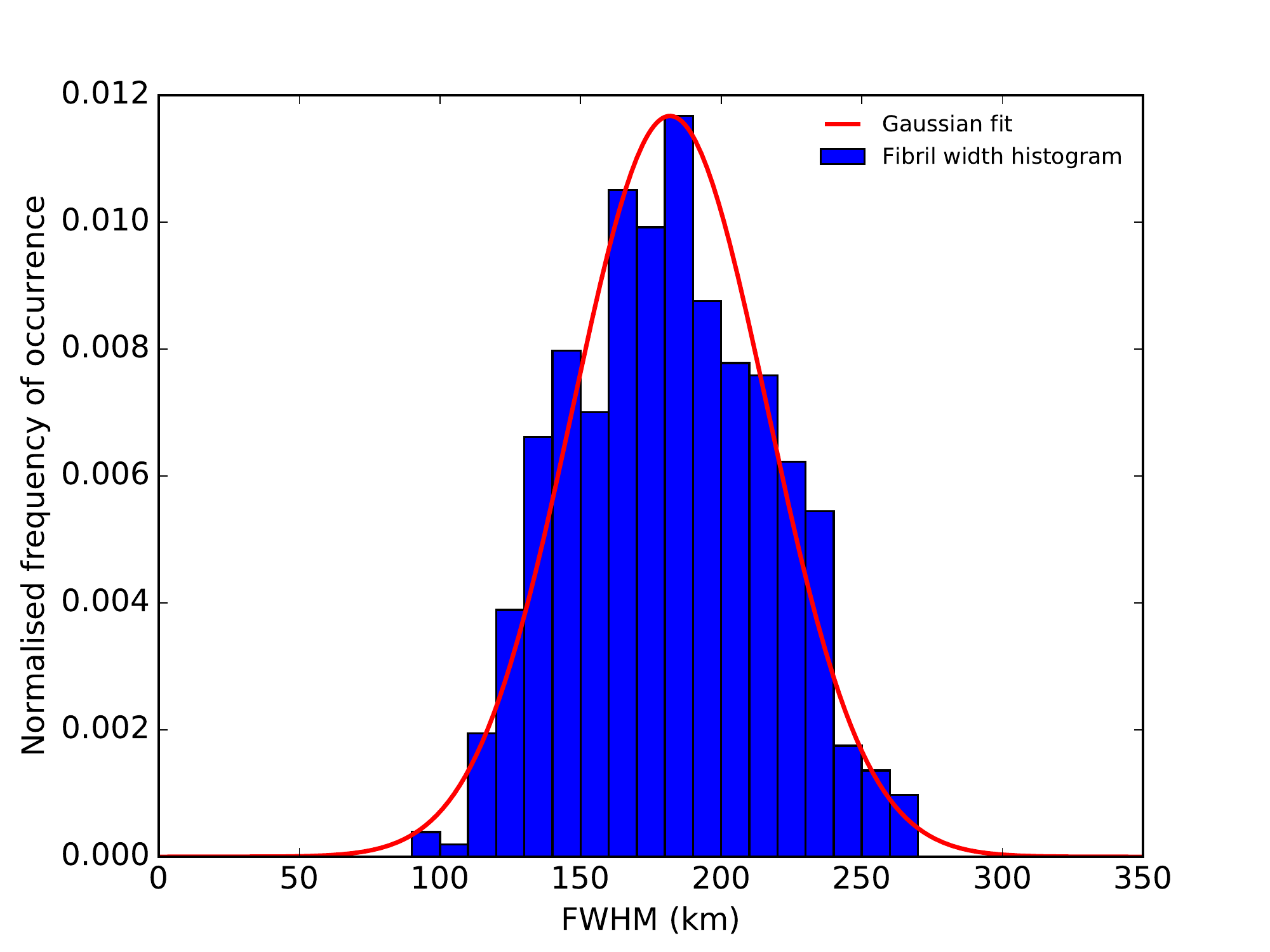}
\caption{Width distribution for each individual fibril detected in all the frames and its respective Gaussian fit (red curve).}
\label{width}
\end{figure}

The above procedure is repeated for all fibrils in each frame to produce the histogram presented in \fig{width}. This  distribution of fibril widths is reasonably well represented by a Gaussian function (red line) with a mean width of $\approx$182\,km (see \tab{morphprop} for the other fit parameters of this Gaussian). The narrowest fibril was determined to have a width slightly smaller than 100\,km, which is above the spatial resolution of \sunrise{} of around 70\,km at the wavelength of \caiih{}. This fact might suggest that $\approx$100\,km is a lower limit for the width of solar fibrils, and that \sunrise{} is actually resolving the fibrils. However, we are aware of the fact that the determination of the fibril width is not without problems: The fibrils have to be identified against a relatively bright photospheric background contamination resulting from the broad transmission profile of the \caiih{} filter used in \sufi{}. Additionally, the width determination method, based on the fitting of a Gaussian function, is likely to smear out structures at the diffraction limit of the telescope, resulting in a larger minimum width. 

The study of the width of \caiik{} fibrils by \citet{Pietarila2009} showed minimum fibril widths of only 70\,km, corresponding to the diffraction limit of the SST. Their analysis method was based on a binary mask, which is possibly better suited to determining the minimum widths of the fibrils, but has disadvantages in properly characterizing broader fibrils. When applying our technique to a sample of the fibrils in the data set used by \citeauthor{Pietarila2009}, we also get $\approx$100\,km for the minimum width of their \caiik{} fibrils, in accordance with our \sunrise/\sufi\ measurements. We therefore conclude that the minimum width of our \caii{} fibrils is likely to be set by the spatial resolution of the \sunrise{} telescope.

\subsection{Length}\label{leng}

An important parameter returned by the automatic tracking method is the length of the detected SCFs, which we define as the length of the reference backbone as described in \sref{detmet}, plus the average of the widths computed in \sref{Widths} at both end points. This addition is necessary to compensate for the effect when reducing the fibril to the single-pixel backbone, which reduces the backbone length on both ends approximately by the width of the fibril. The length of the fibril as determined here would correspond to the distance from end-to-end of the shaded area in each of the right-hand panels of \fig{fibilexample}. 
Fibrils at the edge of the field of view are excluded from the analysis.

\begin{figure}[ht]
 \centering
 \includegraphics[width=\figwidth]{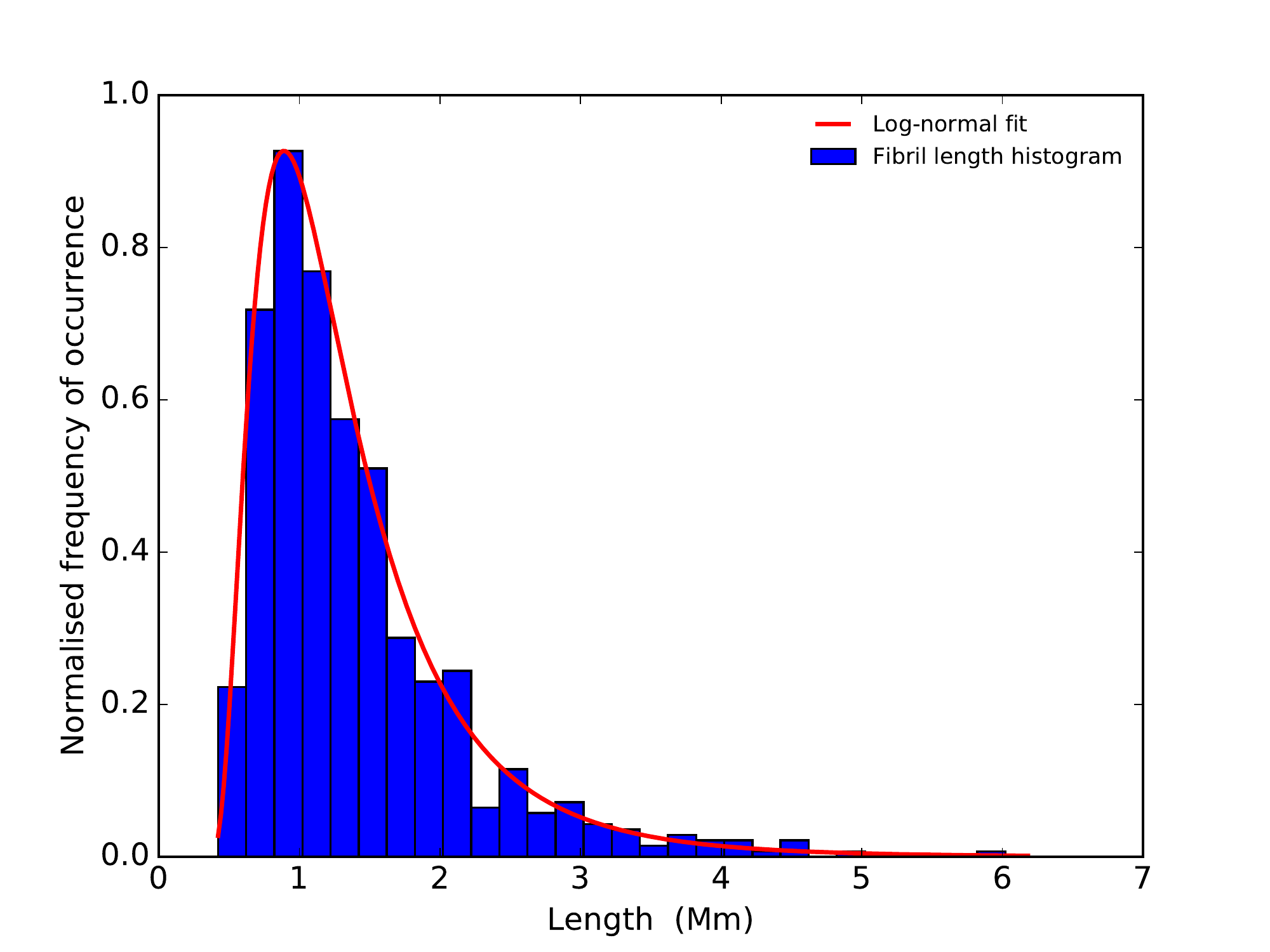}
 \caption{Distribution of the backbone length, averaged over the lifetime of the SCFs (blue), and respective log-normal fit (red curve).
}
 \label{length}
\end{figure}
   
The histogram in \fig{length} shows that the detected fibrils are rather short, with a minimum length of $\approx$500\,km. The mean length of the fibrils is $\approx$1300\,km, and we could not find any fibril longer than 4500\,km.
The length distribution nicely follows a log-normal distribution, indicated by the red line in \fig{length} (for the fit parameters see \tab{morphprop}).

Just as with the longer-lived fibrils, also the longer fibrils are more likely to be excluded because they touch one of the boundaries of the field-of-view. We again follow \cite{Danilovic2010} and multiply the frequency of occurrence of fibrils that have a length of $m$ pixels by the factor $(n-2)/(n-1-m)$, where $n$ is the distance from one edge of the frame to the other in the average direction of the fibril.
This correction has been taken into account in \fig{length}. The average length of the fibrils is found to be  1.38\,Mm. It is worth noting that the length does not correlate very well with the lifetime of the fibrils (the correlation coefficient is only 0.3).

The short length of the fibrils may be an artifact of the automated identification technique, so that we are looking at fragments of fibrils in many cases \citep[see the discussion in][ and in \sref{detmet}]{Pietarila2009}. In the blow-up of the central section of the \caiih\ image plotted in \fig{allims} B (data after MFBD reconstruction) many long fibrils are seen, some of them reaching 10\arcsec{} in length.
However, these long fibrils are often interrupted, or cut by other structures, making it hard for our simple identification algorithm to recognize them as a single long entity. Instead, the identification procedure breaks them into shorter fibrils. 

There is also a general geometrical separation between the longer and shorter fibrils, at least as seen by eye, in the sense that the longer fibrils are found more in the central-lower part of \fig{allims} B directed roughly E-W, while the shorter fibrils are located in the upper part of the figure, in particular surrounding the small pores. In addition, the pattern of seemingly very short fibrils, or rather small bright patches in the lower left part of the image may be produced by crisscrossing longer fibrils. 
   
\subsection{Curvature}\label{curv}

The reference backbone of most of the fibrils can be fitted nicely with a second order polynomial. This has partly to do with the fact that the automatically identified fibrils are rather short, but also because the fibrils are expected to outline loop-like magnetic flux tubes \citep[although rather flat ones; see][]{wiegelmann2016,jafarzadeh16d}: the slanted view at a heliocentric angle of $\theta\approx 22^\circ$ makes loop-like structures appear similar to parabolas.

\begin{figure}[ht]
   \centering
   \includegraphics[width=\figwidth]{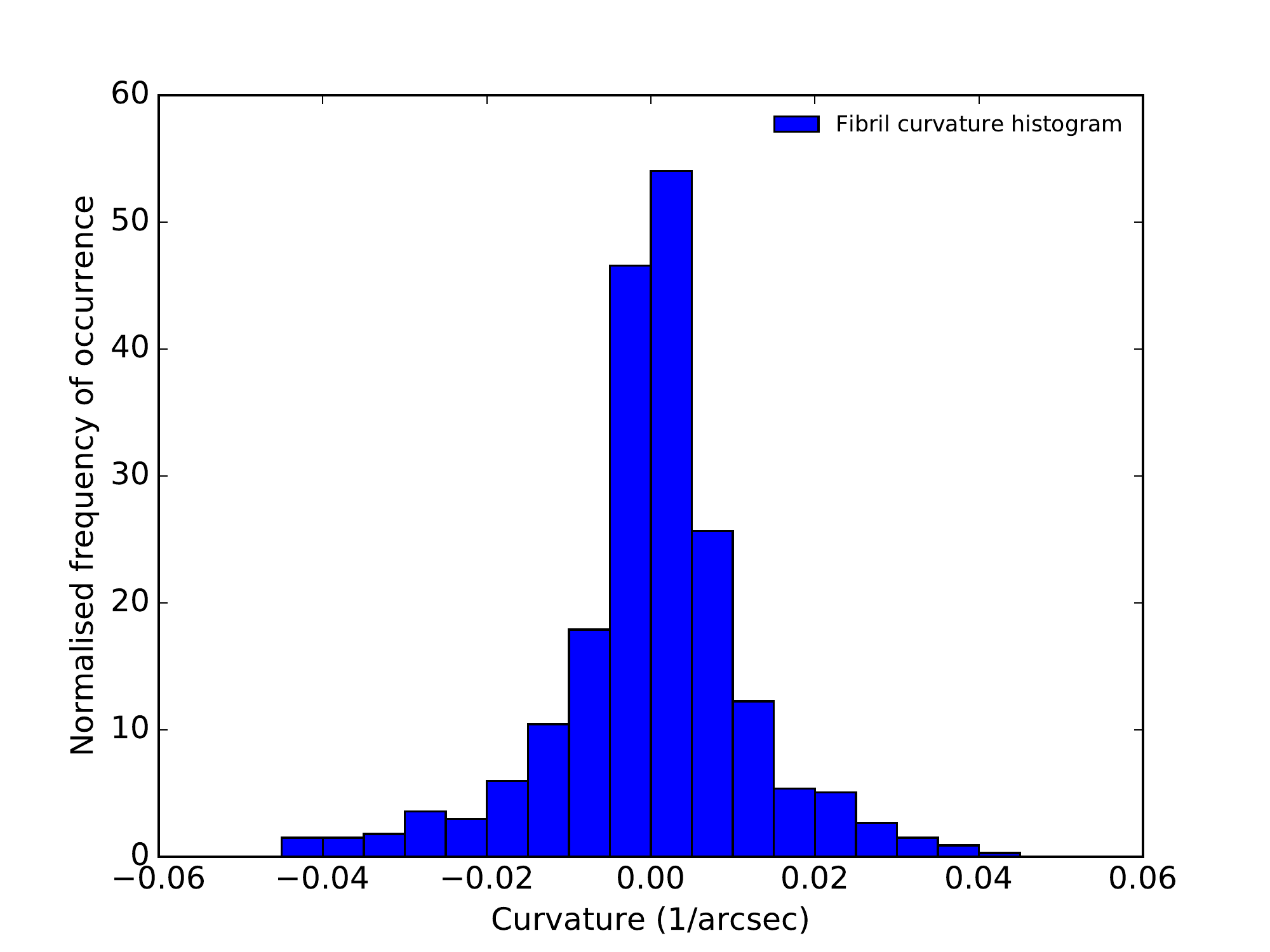}
   \caption{Curvature distribution of the fibril's reference backbones.}
   \label{curvature}
   \end{figure}

We define the curvature, $\kappa$, of a fibril as the reciprocal value of the radius of the largest circle fitting the fibril tangentially, i.e., the so called osculating circle. For a second order polynomial, $\kappa$ is defined as $2a$, where $a$ is the second order polynomial coefficient.
The histogram in \fig{curvature} shows the curvature distribution obtained from all reference backbones of the time series. The distribution is nearly symmetric with a slight offset towards a positive curvature, with a mean value of 0.002 \,arcsec$^{-1}$.

\subsection{Brightness variation}

The definition of a fibril backbone simplifies not only the study of the morphological properties discussed in the previous section, it also allows visualizing the temporal evolution of intensity fluctuations within the fibril itself. This is achieved by putting an artificial mesh onto the reference backbone of the fibril, with the grid lines being parallel and perpendicular to the backbone. The perpendicular grid lines cover a region of $\pm$0.3\arcsec{} around the backbone, the parallel grid lines have the same length and curvature as the backbone itself. More information on the mesh and gridlines, including an illustration, can be found in \cite{gafeira16b}.

\begin{figure}[ht]
\centering
\includegraphics[width=\figwidth]{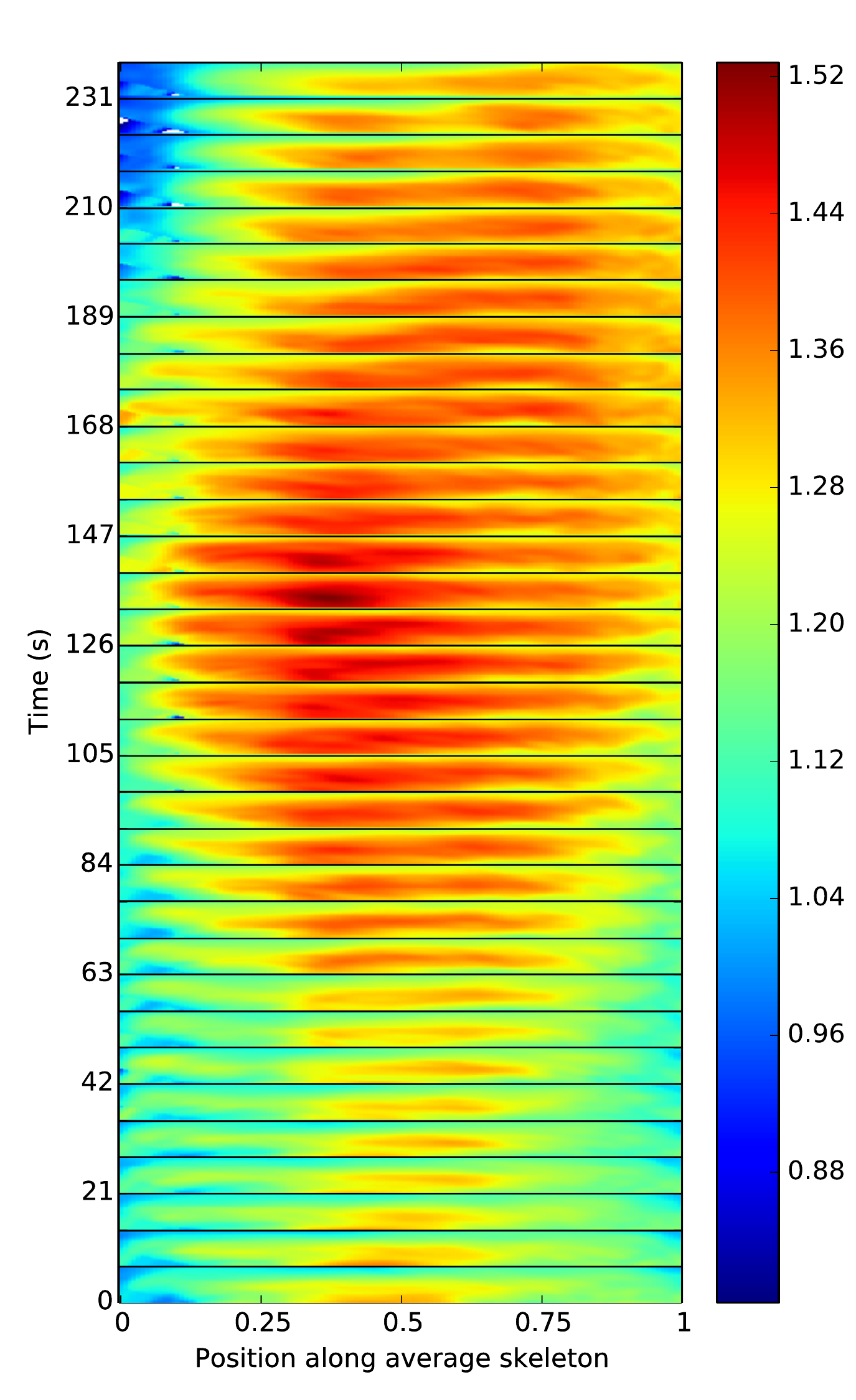}
\caption{Temporal evolution of a SCF with a central brightening. The stacked images show the straightened fibril, with the backbone parallel to the $x$-axis in the center of each bin, surrounded by a $\pm$0.3\arcsec{} wide area covering the fibril. The individual images, recorded every 7\,s, are separated by horizontal black lines. The length of the reference backbone of this fibrils is 1620\,km.
}
\label{centedg}
\end{figure}

\begin{figure}[ht]
\centering
\includegraphics[width=\figwidth]{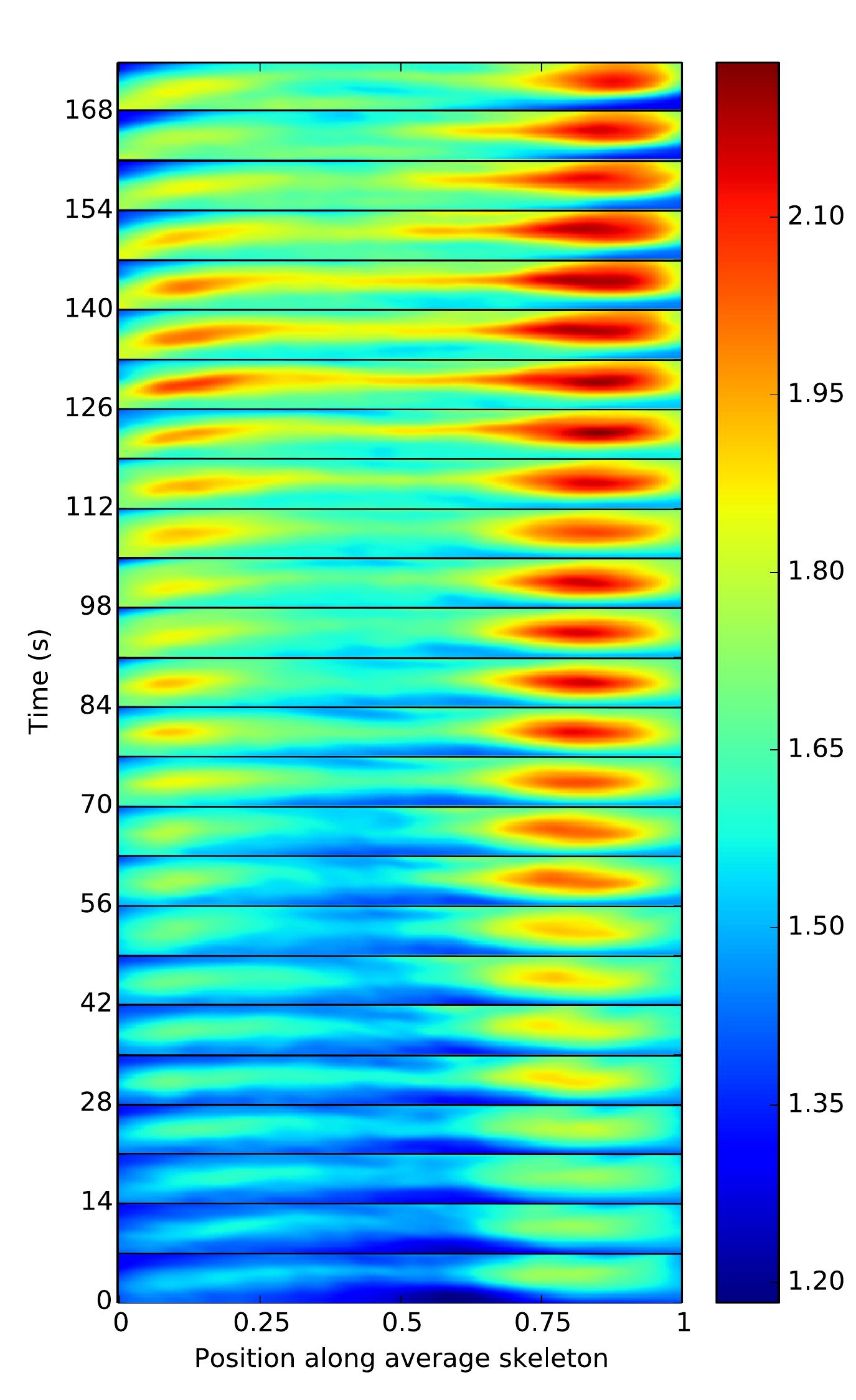}
\caption{Same as \fig{centedg} for a SCF with endpoint brightening. The length of the reference backbone of this fibril is 2800\,km.
\ifdraftold\sks{Same here: give length of image, rather than length of fibril!}\fi}
\label{edgcent}
\end{figure}

This mesh allows to straighten the fibril by interpolating the brightness of the original image to the grid points of the mesh. Since the mesh was defined on the reference backbone of the fibril, we can now nicely follow the temporal brightness variation within the individual fibrils in plots such as the examples presented in \figand{centedg}{edgcent}. These two plots show two different types of brightness variation and propagation along the fibril backbones. In \fig{centedg} the intensity decreases from the center to the endpoints, whereas in \fig{edgcent} the intensity decreases from the endpoints of the fibril to its center. After visual inspection of all fibrils we find that the first type is the more common (around 54\% of the the fibrils are of this type), and the second type is observed in $\approx$25\% of the fibrils. For the remaining $\approx$21\% of the fibrils it is difficult to clearly categorize the distribution of the brightness along the fibril axis, with only a very small fraction of those showing a brightening at one endpoint only. We could not find any obvious correlation between the type of the brightening (central or endpoint) with other parameters of the fibrils (length, width absolute value, lifetime). It can also be seen that the width of the fibrils is not constant along the the length of the fibril. In general, the fibrils are broader at the locations where they are also brighter.

Note that in some cases, a fibril changes from having its brightness peak at its center to having it near its ends. Obviously, the brightness of the fibril varies with time, as can be seen in \figand{centedg}{edgcent}. \cite{gafeira16b} have pointed out that these variations are often periodic and occur mainly in (anti-)phase with oscillations in the width of the fibril. They identified these variations with sausage-mode waves.

\begin{table*}
\centering
\caption{Summary of the morphological properties of the SCFs.}   
\label{morphprop}
\setlength{\extrarowheight}{3pt}
\newlength{\cw}
\setlength{\cw}{2.7cm}
\begin{tabularx}{\linewidth}{L{\cw}C{\cw}C{\cw}C{\cw}C{\cw}C{\cw}} \hline\hline

para\-meter  & lifetime \newline (excl. gaps) & lifetime \newline (incl. gaps) & length & width & curva\-ture\\ \hline
   functional form      & expo\-nential & log-normal & log-normal & gaussian & sym\-metric \\
   range                & 35--450\,s  & 35--2000\,s & 500--4500\,km & 100--270\,km & -0.04--0.04\,arcsec$^{-1}$ \\
   mean value           & n/a & 446\,s & 1380\,km & 182\,km & 0.002\,arcsec$^{-1}$ \\
   standard deviation   & n/a & 310.27 &  760\,km&34\,km & 0.019 \\
   skewness             & n/a & 3.70 &  2.51    &   n/a  & 2.26 \\
   kurtosis             & n/a & 31.60 &  3.24    &   n/a  & 17.83 \\
   exp. decay rate      & 25\,(ms)$^{-1}$ & 3.0\,(ms)$^{-1}$ &   n/a    &   n/a   &    n/a    \\ \hline
\setlength{\extrarowheight}{1pt}
\end{tabularx}
\end{table*}

\section{Discussion and conclusions}\label{discussion}

In this paper we present a first quantitative study of the morphological properties of slender \caiih{} fibrils (SCFs) aimed at measuring their parameters, such as lifetime, length, width, curvature and brightness structure. We apply an automatic fibril detection algorithm based on several unsharp masking steps to identify the backbone of the fibrils (i.e., the curve following the central brightness ridge along the long-axis of the SCF). This allows us to identify, measure properties of and perform a  statistical analysis on 598 SCFs. We obtain values for the lifetime between half a minute and half an hour (when taking into account the intermittency in fibril visibility),
lengths between 0.5 and 4.5\,Mm, and widths between 100 and 270\,km. The fibrils show a wide distribution of curvatures, with a slight preference for a positive curvature, in agreement with the expected curvatures of low-lying loops observed at a position away from the solar disk center. 
We also identify two different morphologies of the evolution of the brightness within individual fibrils. In one type the intensity increases starting from the center and moving to the edges, while in the other fibrils the increase in intensity first becomes visible near one or both of the ends and expands from there towards the center of the fibril.

Intersecting fibrils and not well separated fibrils cause problems in the automatic tracking of the fibrils, mainly leading to a failure in detecting fibrils longer than 4.5\,Mm, albeit the visual inspection of the \sufi{} \caiih{} images shows that such fibrils do exist. 
The contrast between the measured short lengths and the pattern of extended fibrils in parts of the \sufi{} FOV suggests that the automated technique is identifying relatively undisturbed parts of fibrils as whole fibrils. Between these undisturbed fragments a given long fibril may be intersected by other fibrils, or disturbed by underlying bright points, reversed granulation and acoustic grains. This complex picture may arise because at the height of the SCFs, the \caiih\ line may not be entirely optically thick, so that radiation from below can pass through possibly multiple layers of fibrils. This may have to do with the 1.1\,\AA{} FWHM of the filter employed by \sufi{}, which includes not just the H$_2$ and H$_3$ features of the line core,  but also parts of the inner wings. The H$_2$ and H$_3$ features sample the lower to middle chromosphere, while the inner wings are mainly sensitive to the upper photosphere \citep[see heights of formation computed by][]{Danilovic2014,jafarzadeh16a}. 

\begin{table}
\centering
\caption{Comparison of some morphological properties between SCFs observed in \caiih\ and in and \caiik{} spicules.
}
\label{comp}
\begin{minipage}{\linewidth}
\setlength{\extrarowheight}{3pt}
\begin{tabularx}{\linewidth}{p{3cm}p{1.6cm}ccc} \hline\hline
   & lifetime & width & length\\ \hline
  SCFs (this work)& 35--450\,s \newline (35--2000\,s)\footnote{after taking into account the intermittent fibrils, see \sref{sec:lifetime}}  & 100--500\,km & 0.5--4.5\,Mm\\
  \mbox{}\hfill mean value & 80\,s     & 182\,km & 1380\,Mm\\ \hline
  type I spicules\footnotemark[2]& 150--400\,s &100--700\,km& 4--8\,Mm\\
  \mbox{}\hfill mean value & 262\,s     & 348\,km & 6.87\,Mm\\ \hline
  type II spicules\footnote{\cite{Pereira2012}}& 50--150\,s &100--700\,km&3--9\,Mm\\ 
  \mbox{}\hfill mean value & 165\,s     & 319\,km & 7.75\,Mm\\ \hline
  \caiik{} \footnote{\cite{Pietarila2009}} &&75--150\,km& \\ 
  \mbox{}\hfill mean value &      & 100\,km & $\approx$ 0.9\,Mm\\ \hline
\setlength{\extrarowheight}{1pt}
\end{tabularx}
\end{minipage}
\end{table}

The results of the present analysis allow us to relate SCFs to small-scale elongated, chromospheric structures already known from the literature. In \tab{comp} we list the properties of the SCFs together with those of type I and type II spicules \citep[from][]{Pereira2012}, and of \caiik{} fibrils \citep[as deduce by ][]{Pietarila2009}. Spicules show considerable similarities to the SCFs with regard to the range of widths (although the Ca fibrils studied by us and, in particular, those by \citeauthor{Pietarila2009} are on average narrower).

The difference in width between the fibrils seen by \citeauthor{Pietarila2009} in \caiik{} and those studied by us in \caiih{} is puzzling. It has at least partly to do with the difference in technique, as the application of our technique to a sample of fibrils identified by \citet{Pietarila2009} returns larger widths than those found by those authors. Other factors may also contribute to the difference in width. Thus, it may have to do with different optical thicknesses, or that our observations were recorded in a larger, more intensely active region.

Both types of known spicules are considerably longer than our SCFs, with even the shortest spicules being close to the longest SCFs that we find. This difference in length may be a selection effect, however, as only the longer spicules are clearly identified at the limb, whereas shorter ones are buried in the forest of features at the solar limb. Conversely, our technique most likely underestimates the lengths of the fibrils, as it often fails to follow the fibrils to their usually faint ends and often identifies the undisturbed fragments of a fibril (lying, e.g., between locations at which the fibril crosses other structures) as complete, separate fibrils.  
The lifetimes of spicules appear to lie in between those we obtain without considering SCFs to survive gaps and those obtained after interpolating across such gaps. 

Such a similarity in some properties does not imply that SCFs and spicules both occur in similar magnetic structures. Firstly, spicules are clearly identified only at some distance from the limb, i.e., at a much greater height than the SCFs, which are clearly located in the low chromosphere. Secondly, whereas spiculare are more or less outward directed, as clearly shown by limb observations, the SCFs we see are well visible on the disk over a distance of up to $10^4$ km, which implies that they are nearly horizontal. This conclusion is supported by the comparison carried out by \citet{jafarzadeh16b} between the SCFs imaged by \sufi{} and the magnetic field extrapolated into the chromosphere on the basis of a magnetostatic equilibrium \citep{wiegelmann2016}. \citet{jafarzadeh16a} found that the azimuthal directions of both agreed very well, that the magnetic field lines corresponding most closely to the fibrils remained nearly horizontal. These field lines also mostly returned to the solar surface, thus forming very flat, low-lying magnetic loops. However, there are also (shorter) fibrils that are associated with field lines that reach up into the corona. These may be more closely related to spicules. 

Also, we find that brightenings in around 50\%\ of the SCFs  start in the body of the fibril, rather than at its edge. These features are obviously not driven from one footpoint, as are jet-like structures such as spicules. For another $\approx$25\% of the SCFs, the brightness variation is similar to that presented in  \fig{edgcent}, with the two edge-points of the SCF being brighter than the central part, which is also not compatible with a jet-like structure, where only one end of the structure is rooted in the photosphere, from where it points towards the upper layers of the atmosphere. All this suggests that our SCFs are more likely to outline small, loop-like structures, an interpretation which is also supported by the analysis of the curvature computed in the \sref{curv}. The brightness enhancement at both ends of the SCFs may be a result of a footpoint heating process, although the structure may simply reflect the fact that the broad Ca filter of \sufi{} allows the photospheric bright points to shine through. This is very similar to the structures seen by \citet{Pietarila2009}, who also found many of their narrow fibrils to be rooted in underlying bright points.  

The radiation transmitted through the \sufi{} 1.1\,\AA\ \caiih{} filter comes from either the photosphere, or the lower chromosphere (average height of 600--700\,km). This is very close to the limit of the small-scale canopies created by the granulation \citep{Wedemeyer2008}, which are also found by \citet{jafarzadeh16a} on the basis of the MHD-simulation assisted Stokes Inversion (MASI) output of \citet{riethmuller2016}. Some of the shorter SCFs may indeed be associated with the small-scale granular canopies, with possible similarities to the seething horizontal magnetic fields described in \cite{Harvey2007}, or the horizontal fields detected by Hinode \citep{2008ApJ...672.1237L} and \sunrise\ \citep{2010ApJ...723L.149D}. However, the longer SCFs are likely more closely related to larger-scale canopy found over the network by \cite{Jones1982} and modeled by \cite{Solanki1990}. Such extended horizontal field lines were found lying nearly parallel to the SCFs in the magnetostatic equilibrium computed by \cite{wiegelmann2016}.

\begin{acknowledgements}
The German contribution to \sunrise{} and its reflight was funded by the
Max Planck Foundation, the Strategic Innovations Fund of the President of the Max Planck Society (MPG), the Deutsche Zentrum f\"ur Luft- und Raumfahrt (DLR), and private donations by supporting members of the Max Planck Society, which is gratefully acknowledged. The Spanish contribution was funded by the Ministerio de Econom\'{i}a y Competitividad under Projects ESP2013-47349-C6 and ESP2014-56169-C6, partially using European FEDER funds. The HAO contribution was partly funded through NASA grant number NNX13AE95G. This work was partly supported by the BK21 plus program through the National Research Foundation (NRF) funded by the Ministry of Education of Korea.
SJ receives support from the Research Council of Norway.
\end{acknowledgements}

\bibliographystyle{apj}

\end{document}